\documentclass[aps,preprint]{revtex4}%
\usepackage{amsmath,amsfonts,amssymb,amsopn}
\usepackage{graphicx}%
\begin{document}
\title{New longitudinal mode and compression of pair ions in plasma}

\author{Zahida Ehsan${}^{1}$, N. L. Tsintsadze${}^{2}$, H. A. Shah${}^{3}$, R. M. G. M. Trines ${}^{4,5}$ and M. Imran ${}^{1^*}$}

 \affiliation{
$^{1}$ Department of Physics, COMSATS Institute of Information Technology, Lahore 86090, Pakistan.\\
$^{2}${Faculty of Exact and Natural Sciences and Andronicashvili Institute  of Physics, Javakhishvili Tbilisi University,Tbilisi 0128, Georgia.}
$^{3}$ GC University Lahore, 54000, Pakistan.
$^{4}$Central Laser Facility, STFC Rutherford Appleton Laboratory, Didcot, OX11 0QX, UK\\
$^{5}$Department of Physics, Lancaster University, Lancaster, LA1 4YW, UK.
\footnote{For correspondence: imransindhu@hotmail.com}}

\date{\today}

\begin{abstract}

Positive and negative ions forming so-called pair plasma differing in sign of their charge but asymmetric in mass and temperature support a new acoustic-ike mode. The condition for  the excitation of ion sound wave through electron beam induced Cherenkov instability is also investigated. This beam can generate a perturbation in the pair ion plasmas in the presence of electrons when there is number density, temperature and  mass difference in the two species of ions.
 Basic emphasis is on the focusing of ion sound waves and we show how, in the area of localization of wave energy, the density of pair particles increases while electrons are pushed away from that region. Further, this localization of wave is dependent on the shape of the pulse. Considering the example of pancake and bullet shaped pulses, we find that only the former leads to compression of pair ions in the supersonic regime of the focusing region. Here  possible existence of regions where pure pair particles can exist  may also be speculated which is not only useful from academic point of view but also to mimic the situation of plasma (electron positron asymmetric and symmetric ) observed in astrophysical environment.

\end{abstract}
\maketitle

\section{Introduction}

Pair (pi) plasmas consisting of only positive and negative ions, produced in laboratory, have received considerate\cite{1, 2, 3, 4, 5, 6, 7, 8, 9, 10, 11} attention for the reason that such plasmas have allowed the investigation of many classic problems of physics by opening a door for the scientists curious to understand the phenomena occurring hundred of light years  away in the astrophysical environment, this motivates us to address this an advancement in the ``laboratory astrophysics''. 

Electron and positron (pair) plasmas have been intensively studied in the preceding decades. The latter are found in some interstellar compact objects (e.g., in neutron stars, in the interior of Jupiter, in active galactic nuclei (AGN), etc.), near the polar cusp regions of the pulsars and neutron star atmospheres, in the inner region of the accretion disks surrounding the central black hole, in quasar atmospheres, and in the Van Allen radiation belts etc. \cite{12,13,14,15,16,17,18}.  It is also known that the early prestellar period of the evolution of the Universe was presumably dominated by relativistic electrons and positrons \cite{19}. 

In the lepton epoch, which occurred $10^{-6} < t < 10$ sec after the Big Bang, temperatures reached the values of $10^{9} < T < 10^{13}$ K causing annihilation of the nucleon-antinucleon pairs resulting in matter composed of the electrons, positrons, and photons in thermodynamic equilibrium \cite{19}. 
Unlike ordinary plasmas with significant differences between the masses of chased particles, pair plasmas on the other hand keep a time-space parity because the mobility of the particles in electromagnetic fields is the same.  Thus the symmetry collapses the scales which disentangle short and long wavelengths.  Positrons have been focused in connection with antimatter property, e.g., CPT (charge, parity and time reversal) invariance, in high- energy physics and astrophysics. Both the relativistic and non-relativistic pair plasmas have been gradually explored to represent a new state of matter with unique thermodynamic properties drastically different from ordinary electron-ion plasmas. For these reasons it's very challenging to produce a pair plasma in the laboratory. However,  successful creation of sufficiently dense pair plasma made of fullerene 
ions $C_{+}^{60}$ and $C_{- }^{60}$ is indeed a triumph of experimental science, Oohara
and Hatakeyama\cite{1, 2} have intrigued the scientific community. 
This is because in such plasmas we do not encounter with the annihilation problem which
arises (at much longer characteristic time scales compared to the 
collective interaction time) due to interaction of matter (e.g., electrons) 
and anti-matter (positrons) and so provide a controlled environment to study the underlying physics. A basic requirement for long time 
scale experiments will be that the pair annihilation time scale is many orders of magnitude 
larger than the plasma period.
However, the  disagreement between experimental observations and basic fluid theory is yet to be resolved and has divided the concerned scientific community \cite{4, 5, 6, 7}.


Pair plasmas have been dealt with two different ways: One is recognized as a ``symmetric''
system where pair particles have the same charge, mass, temperature, 
density etc., whereas in second way of treatment, the symmetry 
of the pair plasma is mildly broken and the system is usually known as ``asymmetric'' plasma.
   
This asymmetry, however,  brings forth new physics frontiers, as is of interest as such plasmas can be produced in the laboratory. Whereas some nonlinear phenomena which emerge naturally during the evolution of pair particles may usually cause this asymmetric behavior in the experiments. Small temperature differences in the constituent species causing asymmetries can lead to interesting nonlinear structure formation in astrophysical settings where one encounters e-p plasmas and in laboratory produced pair ion plasma, whereas in the latter small contamination by a much heavier immobile ion, or a small mass difference between the two constituent species can also produce asymmetries \cite{20, 21, 22, 23}. 
 
In Japan, Hatekayama and Oohara \cite{24, 25} succeeded in creating lighter pair plasma with hydrogen; however, efforts are being made to accomplish its improved quality. In parallel, for the theoreticians it's a challenge to explain some of the results and some attempts have been made with kinetic theory taking into account the boundary effects \cite{7}. 
 
Looking at the results presented  by  Oohara et al., \cite{1,2}, some authors pointed out that the produced pair-ion fullerene plasmas seem to contain electrons as well since the ion acoustic wave observed in experiment cannot be observed in a pure a pair ion plasma at the same temperature \cite{6}. Later on, criteria to define pure pair ion plasma was also presented and it was shown that the electrons are not fully filtered out and the observation of one of the linear modes proves their presence in the system . And that the increase
 in the concentration of electrons in pair-ion plasmas  affects the speed of ion acoustic wave (IAW) corresponding to the same electron temperature\cite{4}. 
 
Verheest et al. \cite{5} demonstrated that a strict symmetry destroys the stationary nonlinear structures of acoustic nature and showed such nonlinear structures can exist when there is a thermodynamic asymmetry between both constituents.

In this manuscript, we aim to investigate linear longitudinal mode for asymmetric pair plasma and  with an additional concentration of electrons. We note that not only asymmetry leads to a new longitudinal mode but also the addition of electrons is useful in studying the pure pair ions. Nonetheless we discuss few cases with and without concentration of electrons. It's to be noted that our investigations are strictly valid for non-relativistic case.   

The organization of the paper is as follows: In Sec. II, we discuss the dispersion relation for the cases without and with the presence of electrons. In Sec. III, Cherenokov instability, and the condition for the excitation of the ion sound wave is studied. Sec. V devotes to the derivation of Zakharov equations with and without concentration of electrons. The discussion of the compression phenomenon and the possibility of formation of pair plasma is presented in Sec. VI.  The main findings of the paper are recapitulated in Sec. VII.

\section{Longitudinal wave}

It has been reported by \cite{1,2} that a pure pair-ion plasma can support three kinds of electrostatic ion waves in 
different frequency regimes propagating parallel to the external static magnetic field. 
Where the ion acoustic speed has been defined as $c_{s}= {T_{i}/m_{i}}$ where ${m_{i}}={m_{+}}={m_{-}}$ and ${T_{i}}={T_{+}}={T_{-}}$ have been assumed. The subscripts plus and minus denote the singly charged positive and negative ions. This is also recognized as the case of a symmetric pair ion plasma.
Observing the dispersion properties in fullerene pair plasma experiments, Verheest et al., emphasized that there seem to exist an acoustic wave rather than linear electrostatic waves interpreted by the Ohara et al. \cite{2} which emphasizes symmetry breaking. This is also consistent with the experimental conditions pointing to small asymmetries and is point of our interest. For this let's  start our analysis for the spectrum of linear wave associated with the pair plasma, for that we first write Poisson's equation for all species:
 
\begin{equation}
\epsilon k^2\phi =-e(n_{e}+Z_{-}n_{i-}-Z_{+}n_{i+})  \tag{1}
\end{equation}

 where $Z_{\pm}$ is the charge on positive and negative ions respectively and
$n_{\alpha0}$ is the number density with $\alpha=e$ (electrons), $i+\;$(positive ions)
and $i-$ (negative ions). The equilibrium quantities will be written with
subscript ``0''.

For the longitudinal electrostatic perturbations $~\exp {-i}\left({\omega {t}} -{kx}\right)$, we obtain following dispersion relation:

\begin{equation}
\epsilon(\omega,k)=\frac{1}{k^{2}\lambda_{D{\alpha}}^{2}}\left[ 1- W(\widetilde{\xi }_{\alpha})  \right]  + {. . .}  \tag{2}
\end{equation}

where $\widetilde{\xi }_{\alpha }={\omega }/{k_{z}v_{t\alpha }}$ and
\begin{equation}
W(\widetilde{\xi }_{\alpha})=1/\sqrt{ 2\pi }\int \limits_{-\infty }^{\infty }
 x \frac{e^{-x^{2}/2}}{x-\widetilde{\xi }_{\alpha }}dx,
\tag{3}
\end{equation}

is the plasma dispersion function  and asymptotic expressions of which are:

\begin{equation}
W(\widetilde{\xi }_{\alpha })=1+\frac{1}{\widetilde{\xi }_{\alpha }^{2}}+
\frac{3}{\widetilde{\xi }_{\alpha }4}+...-i\sqrt{\frac{\pi }{2}}\widetilde{
\xi }_{\alpha }e^{-\frac{\widetilde{\xi }_{\alpha }^{2}}{2}\text{ }}\text{for
}\left \vert \mathop{Re}\widetilde{\xi }_{\alpha }\right \vert >>\left \vert 
\text{Im}\widetilde{\xi }_{\alpha }\right \vert 
 \tag{4}
\end{equation}

\begin{equation}
W(\widetilde{\xi }_{\alpha })=-i\sqrt{\frac{\alpha }{2}}\widetilde{\xi }
_{\alpha }\text{ for}\left \vert \widetilde{\xi }_{\alpha }\right\vert \ll 1
\tag{5}
\end{equation}

where $\lambda _{D\alpha }=(\epsilon K_{B}T_{\alpha }/n_{0\alpha }e^{2})^{1/2}$ represents the Debye length, $\omega$, and $k$ are the frequency and propagation vector of the perturbations. 
Different species, not produced in identical conditions, for instance, could have different thermal speeds
temperatures. One could also arrange experiments with different setups for
different species when, for instance, there are fractions of heavier or
lighter ions or there is a mixture of different mass or temperature species
with opposite charges. Below we discuss dispersion  of the waves for different cases.  

{\bf{Case 1}}:\bigskip 

First we consider a case when there is temperature and mass asymmetry in a two component pi plasma, for this it is assumed that $T_{+}$ is the temperature of lighter ions with mass $m_{+}$.
\begin{equation}
\epsilon (\omega, k)=1+\frac{1}{k^{2}\lambda _{D+}^{2}}\left[  1- W(\widetilde{\xi }_{+})   \right] +\frac{1}{k^{2}\lambda _{D-}^{2}}
\left[1- W(\widetilde{\xi }_{-})\right] 
\tag{6}
\end{equation} 

For the frequency range  $v_{t-} < \omega/{k} < v_{t+}$, we obtain
 
\begin{equation}
1+\frac{1}{k^{2}\lambda_{D+}^{2}}\left(  1+i\sqrt{\frac{\pi}{2}}\frac{\omega
}{kv_{t+}}\right)-\frac{\omega_{p+}^{2}}{\omega^{2}} -\frac{\omega_{p-}^{2}}{\omega^{2}}+\sqrt{\frac{\pi }{2}}\frac{1}{k^{2}\lambda _{D-}^{2}}\frac{\omega}{kv_{t-}}\exp \left( -\frac{\omega^{2}}{2k^{2}v_{t-}^{2}}\right) =0
\tag{7}
\end{equation}

where $v_{t\pm }=(K_{B}T_{\pm }/m_{\pm})$ is the thermal velocity of positive and negative ions. 
The damping term of ions is exponentially small. If we put $\omega=\omega_{r} + \omega_{i}$ in above equation,
 then we obtain the real ($\omega_{r}$) and imaginary part
($\omega_{i}$) of frequencies as:

\begin{equation}
\omega _{r}=\frac{k\left( \frac{T_{+}}{m_{-}}\right) ^{1/2}}{\left(
1+k^{2}\lambda _{D+}^{2}\right) ^{1/2}}
\tag{8}
\end{equation}

and
\begin{equation}
\omega _{i}=-\sqrt{\frac{\pi }{8}}\frac{k\left( \frac{T_{+}}{m_{-}}\right)
^{1/2}\left( \frac{m_{+}}{m_{-}}\right) ^{1/2}}{\left( 1+k^{2}\lambda
_{D+}^{2}\right)^{2} }\left[ 1+\left( \frac{T_{+}}{T_{-}}\right) ^{3/2}\left( 
\frac{m_{-}}{m_{+}}\right) ^{1/2}\exp \left( -\frac{1}{2}\frac{\left( \frac{
T_{+}}{T_{-}}\right) }{(1+k^{2}\lambda _{D+}^{2})}\right) \right] 
 \tag{9}
\end{equation}

 Eq. (9) is general, where $ \omega_{\pm}= \left(4\pi n_{\pm}e^{2}/m_{\pm}
\right)^{1/2}$. For the damping of light ions $\omega_{i}<\omega_{r}$ but for heavy  ions damping rate is exponential and is
very important when $T_{+}>>T_{-}$.
The mode can exist only for an asymmetric (non isothermal) pair plasma when $T_{+}>>T_{-}$ and $k^{2}\lambda _{D+}^{2}\ll 1$. Since the frequency spectrum depends considerably on the heavier ions so we may call this heavy ion or low frequency branch of the longitudinal oscillations in the pair plasma. When $k^{2}\lambda _{D+}^{2}\ll 1$, the spectrum (8) takes the simple form of the usual ion acoustic oscillations.  
For the condition $|\omega _{i}|\ll \omega _{r}$, we get a limit for the temperature ration of the plasma components.        

\begin{equation}
\left(\frac{T_{+}}{T_{-}}\right) ^{3/2}\exp \left(-1/2\frac{\left( \frac{
T_{+}}{T_{-}}\right) }{(1+k^{2}\lambda _{D+}^{2})}\right) \ll 1 
\tag{10}
\end{equation}

The spectrum (8) and the damping rate (9), we call them the ion sound waves owing to their striking
similarity with the usual ion acoustic waves in an ordinary electron-ion plasma. Whereas this mode can exist only for an asymmetric plasma with $k^{2}\lambda _{D+}^{2}\ll 1$. 
We use some parameters typical for the experiment, the density such as: $%
n_{0}=1\times 10^{8}$ cm$^{-3}$ of the fullerene plasma, the temperature: $%
T_{+}\sim 0.3$eV (in the experiment $T_{+}\sim 0.3-0.5$ eV), thus, $\lambda
_{D+}=0.04$ cm and $v_{t}=200$ m/s. $(k\lambda _{D+})^{2}=0.01,$ then $k=2.5$%
cm, $\omega _{r}=49.75$ kHz or $\omega _{r}/2\pi =5$ kHz, $(k\lambda _{D+})^{2}=0.1$ then $k=7.9$cm, $\omega _{r}=150.647$ kHz, or $%
\omega _{r}/2\pi =47$ KHz.

{\bf{Case 2}}:\bigskip 

 As was pointed out by Saleem et al. \cite{4} that the larger value of the ion acoustic frequency presented in Fig. 2 of Ref.\cite{2}  indicated the presence of electrons in significant concentration. So here we discuss another case, when electrons are also present in the pair plasma. For this $T_{e}>T_{-},T_{+};$ $m_{+}<m_{-};T_{-}<T_{+}$

From special functions given by (2), the real and imaginary parts are given as:
\begin{equation}
\omega _{r}=\frac{k\left[ \left( \frac{T_{e}}{m_{+}}\right) \left( \frac{
Z_{+}n_{+0}}{n_{e0}}\right) +\left( \frac{Z_{-}n_{-0}}{n_{e0}}\right) \left( 
\frac{T_{e}}{m_{-}}\right) \right] ^{1/2}}{\left( 1+k^{2}\lambda
_{De}^{2}\right) ^{1/2}}
\tag{11}
\end{equation}

and 

\begin{equation}
\begin{split}
\omega _{i}=-\sqrt{\frac{\pi }{8}}k\left( \frac{m_{e}}{T_{e}}\right) ^{1/2}%
\frac{\left[ \left( \frac{T_{e}}{m_{+}}\right) \left( \frac{Z_{+}n_{+0}}{%
n_{e0}}\right) +\left( \frac{Z_{-}n_{-0}}{n_{e0}}\right) \left( \frac{T_{e}}{%
m_{-}}\right) \right] }{\left( 1+k^{2}\lambda _{De}^{2}\right)
  ^{2}}\times \\
\times\left[ 
1+\left( \frac{T_{e}}{T_{-}}\right) ^{3/2}\left(
\frac{m_{-}}{m_{e}}\right) ^{1/2} \frac{Z_{-}n_{-0}}{n_{e0}}%
 e^{-\frac{T_{e}}{T_{-}}X}
+\left( \frac{T_{e}}{T_{+}}\right) ^{3/2}
\left( \frac{m_{+}}{m_{e}}\right) ^{1/2} \frac{Z_{+}n_{+0}}{n_{e0}}%
e^{-\frac{T_{e}}{T_{+}}X}%
\right] 
\end{split}
\tag{12}
\end{equation}

where $X=\left[ \left( Z_{+}n_{+0}/n_{e0}\right) \left( m_{-}/m_{+}\right)
+\left( Z_{-}n_{-0}/n_{e0}\right) \right] /2(1+k^{2}\lambda _{De}^{2})$

\section{Cherenkov Instability}

In this section we investigate a situation when a charged particle beam consisting of electrons interacts with 
a pair ion plasma. This situation is widely used in practical application, is a plasma with a small group of electrons of sufficiently high  velocity moving through a medium of ``resting'' particles. The beam density is assumed smaller than that of the plasma. 

We begin our analysis of the plasma-beam system with a straight mono energetic electron beam having a Maxwellian distribution with non relativistic temperature in the intrinsic frame of its electrons penetrating a cold  Maxwellian distributed plasma, however in the laboratory frame. 
It is to be noted that this model is limited to the fast processes with characteristic velocities greatly exceeding the thermal velocities of the beam and plasma particles which, consistently, may be completely ignored. 

 For this the charge neutrality condition alters as: $Z_{+}n_{0i+}\simeq n_{0e}+Z_{-}n_{0i-}+n_{B0}$,  here $n_{B0}$, here $n_{B}$ represents the beam number density. From now on all quantities with a subscript ``B'' will represent the beam electrons. 
 For this case, we obtain the following dispersion relation:

\begin{equation}
\begin{split}
\epsilon(\omega,k)=\ &\frac{1}{k^{2}\lambda_{De}^{2}}\left[ 1- W(\widetilde{\xi }_{e})  \right]  +\frac{1}{k^{2}\lambda_{D+}^{2}
}\left[ 1- W(\widetilde{\xi }_{+})  \right] \\
&+ \frac{1}{k^{2}\lambda_{D-}^{2}}\left[  1- W(\widetilde{\xi }_{-})   \right] + \frac{1}{k^{2}\lambda_{DB}^{2}}\left[ 1-W(\widetilde{\xi }_{b})\right]=0 
\end{split}
\tag{13}
\end{equation}

where $\widetilde{\xi }_{b}={\omega -kv_{0}}/{k_{z}v_{tB}}$ and  $v_{0}$ is the streaming velocity of the beam. Using (4), (5) and (13), we obtain:

\begin{equation}
1+\frac{1}{k^{2}\lambda_{De}^{2}}-\frac{\omega_{p+}^{2}}{\omega^{2}}%
-\frac{\omega_{p-}^{2}}{\omega^{2}}-\frac{\omega_{pB}^{2}}{\left(
\omega-\mathbf{k}\cdot \mathbf{v}_{0}\right)  ^{2}}=0
\tag{14}
\end{equation}

The last term of this dispersion equation is the contribution of the electron beam, which has a second order pole at $\omega \simeq \mathbf{kv}_{0}$. 
When this equality is satisfied, the ``Cherenkov resonance'' takes place. Resonance width is determined by the factor ${1}/{\left(
\omega-\mathbf{k}\cdot \mathbf{v}_{0}\right)}$. For a hydrodynamic instability, the spread in velocities in the beam is much smaller
than the resonance just width. So either all beam particles contribute coherently to the instability, or none of them do.

For $\omega =\omega _{r}+\gamma ,$ where $\omega _{r}$ is given by the Eq.
(11), and 

\begin{equation}
\frac{1}{\omega^{2}}\thickapprox \frac{1}{\omega_{r}^{2}}\left(  1-\frac
{2\gamma}{\omega_{r}}\right)
\tag{15}
\end{equation}

and

\begin{equation}
\frac{1}{\left(  \omega-\mathbf{kv}_{0}\right)  ^{2}}\thickapprox \frac
{1}{\gamma^{2}}\tag{16}
\end{equation}

 and Eq. (14) in terms of $\gamma $ can be written as

\begin{equation}
\gamma ^{3}=\frac{k^{3}\left( \frac{n_{+0}}{n_{e0}}\right) ^{1/2}\left( 
\frac{T_{e}}{m_{+}}\right) ^{3/2}\left( 1+\frac{m_{+}}{m_{-}}\frac{n_{-0}}{%
n_{+0}}\right) ^{1/2}}{2(1+k^{2}\lambda _{De}^{2})^{3/2}}
\tag{17}
\end{equation}

and 
\begin{equation}
\operatorname{Im}\gamma =\frac{\sqrt{3}}{2}\left[ \frac{\omega _{r}\omega _{pB}^{2}}{%
2\left( 1+\frac{1}{1+k^{2}\lambda _{De}^{2}}\right) }\right] ^{1/3}
\tag{18}
\end{equation}

The growth rate (18) describes the hydrodynamic instability in the long
wavelength limit. 

For a kinetic instability, the spread in velocities in the beam is much larger than the resonance width. So only a certain 
fraction of particles having speeds near $\omega/k$ will contribute to the instability (will be in resonance with the plasma wave). 
A phenomenon like Landau damping falls in this class, only it is not normally
seen as an instability since it damps the wave rather than driving it.

Further, we consider the frequency range $v_{te}%
>\omega/k>v_{t+,-}$ for the case of hot electron beam, i.e.,$\left \vert
\left(  \omega-\mathbf{kv}_{0}\right)  \right \vert \ll kv_{tB}$ is satisfied.
In this case, the dispersion Eq. (13)  gives:

\begin{equation}
\begin{split}
1+\frac{1}{k^{2}\lambda_{De}^{2}}\left(  1+i\sqrt{\frac{\pi}{2}}\frac{\omega
}{kv_{te}}\right) +
\frac{1}{k^{2}\lambda_{DB}^{2}}\left(  1+i\sqrt{\frac{\pi}{2}%
}\frac{\left(\omega-\mathbf{k}\cdot \mathbf{v}_{0}\right)  }{kv_{tB}}\right)\\
-\sum_{\pm}\left[\frac{\omega_{p\pm}^{2}}{\omega^{2}}-
\sqrt{\frac{\pi }{2}}\frac{1}{k^{2}\lambda _{D\pm}^{2}}\frac{\omega}{kv_{t\pm}}\exp \left(-\frac{\omega^{2}}{2k^{2}v_{t\pm}^{2}}\right)\right]=0
\end{split}
\tag{19}
\end{equation}

For the case $n_{e0}T_{B}\gg n_{B0}T_{e}$, the real $\omega =\omega _{r}$
coincides with the expression (11). The real and imaginary parts of the
frequency are given as:

\begin{equation}
\omega _{r}=\frac{k\left( \frac{T_{e}}{m_{+}}\frac{n_{+0}}{n_{e0}}+\frac{%
T_{e}}{m_{-}}\frac{n_{-0}}{n_{e0}}\right) ^{1/2}}{\left( 1+\frac{T_{e}}{%
n_{e0}}\frac{n_{B0}}{T_{B}}+k^{2}\lambda _{de}^{2}\right) ^{1/2}}
\tag{20}
\end{equation}

and 

\begin{equation}
\omega _{i}=-\sqrt{\frac{\pi }{8}}\frac{\left( \frac{m_{e}}{m_{+}}\frac{%
n_{+0}}{n_{e0}}+\frac{m_{e}}{m_{-}}\frac{n_{-0}}{n_{e0}}\right) ^{1/2}}{%
\left( 1+\frac{T_{e}}{T_{B}}\frac{n_{B0}}{n_{e0}}+k^{2}\lambda
_{de}^{2}\right) ^{3/2}}\left( \omega _{r}\left(1+ Y\right) +\left( \frac{n_{B0}}{n_{0e}}%
\right) \left( \frac{T_{e}}{T_{B}}\right) ^{3/2}\left( \omega
_{r}-kv_{0}\right) \right) 
\tag{21}
\end{equation}

where 

\[
Y = \left( \frac{T_{e}}{T_{+}}\right) ^{3/2}\left( \frac{Z_{+}n_{+0}}{n_{e0}}%
\right) \left( \frac{m_{+}}{m_{e}}\right) ^{1/2}  \exp \left( -\frac{1}{2}\left( \frac{T_{e}}{T+}\right) \frac{\left[
\left( \frac{Z_{+}n_{+0}}{n_{e0}}\right) \left( \frac{m+}{m-}\right) +\left( 
\frac{Z_{-}n_{-0}}{n_{e0}}\right) \right] }{\left( 1+\frac{T_{e}}{
n_{e0}}\frac{n_{B0}}{T_{e}}+k^{2}\lambda _{de}^{2}\right) ^{1/2}} \right)
\]

In Eq. (21), the first term on right-hand side describes the Landau damping
decrement, the second term can change sign, if we ignore the damping of ions ($Y=0$), $v_{0}>\omega /k\cos \theta $
and $\omega _{i}$ becomes positive when

\begin{equation}
v_{0}>\frac{\omega _{r}}{k\cos \theta }\left[ \omega _{r}+\left( \frac{n_{B0}}{n_{0e}}%
\right) \left( \frac{T_{e}}{T_{B}}\right) ^{3/2}\right] 
\tag{22}
\end{equation}

(22) leads to the kinetic instability. 

\section{Excitation of ion-acoustic wave}

One of the most prominent models in plasma physics is described by the
Zakharov equations\cite{26, 27}, in which high frequency Langmuir waves are coupled nonlinearly to low frequency ion-acoustic waves.
In order to discuss the nonlinear effects in the pair plasma in the hydrodynamic 
approximation, we derive ZakharovÕs equation for both cases such as with and without electrons.
In our consideration, we suppose that high frequency waves are the ion sound waves, it is important to understand that 
when the sound wave interacts with the plasma, it
will strongly perturb the density of {the} plasma. This perturbation is due to
the nonlinearity of {the} equations. Such interactions lead to an average
force which is called the \textit{ponderomotive} force.

{\bf Case 1: Pure pair plasma without electrons.} 
For $T_{+} >T_{-}$, on slow time scale pair ions become dynamic and their equation of motions can be written as:

\begin{equation}
m_{i\pm}n_{i\pm}\left(  \frac{\partial}{\partial t}+\mathbf{v}_{i\pm}%
\cdot \mathbf{\nabla}\right)  \mathbf{v}_{i\pm}=\pm Z_{\pm}en_{i\pm}\left(
\mathbf{E}+\frac{1}{c}\mathbf{v}_{i\pm}\times \mathbf{B}_{0}\right) 
\tag{23}
\end{equation}

and the continuity equation:

\begin{equation}
\frac{\partial n_{i\pm}}{\partial t}+\mathbf{\nabla \cdot}\left(  n_{i\pm
}\mathbf{v}_{i\pm}\right)  =0. 
\tag{24}
\end{equation}

In this case, positive ions being lighter in mass will be effected by the ponderomotive force, so their distribution is given as:  

\begin{equation}
n_{+}=n_{+0}\exp \left( \frac{- \left(e\phi +\frac{e^{2}\left \vert A\right \vert ^{2}%
}{2m_{+}c^{2}}\right)}{T_{+}}\right)   \label{p2}
\tag{25}
\end{equation}

where $A$ is the vector potential of the sound wave and it can be
written as $\mathbf{A}=\mathbf{A}_{0}(r,t)\exp[i(k_{0}\cdot r-\omega_{0}t) \label{vector}$, here $A_{0}(r,t)$ is the amplitude of the vector potential of the
sound wave and it varies slowly with time.

Using the condition $|e\phi +{e^{2}\left \vert A\right \vert ^{2}
}/{2m_{+}c^{2}}| \ll  {T_{+}}$ in (25)  and along with the condition of quasi-neutrality, the potential energy $e\phi$ 
through the density of ions and ponderomotive potential, can be expressed as 

\begin{equation}
e\phi =-T_{+}\left( \frac{Z_{-}\delta n_{-}}{Z_{+}n_{+0}}\right) -\frac{%
e^{2}\left \vert A\right \vert ^{2}}{2m_{+}c^{2}}
\tag{26}
\end{equation}

Substitution of (26) into (23) gives
\begin{equation}
\frac{\partial \mathbf{v}_{-}}{\partial t}+\left( \mathbf{v}_{-}\cdot 
\mathbf{\nabla }\right) \mathbf{v}_{-}=\frac{T_{e}Z_{-}}{m_{-}}\mathbf{%
\nabla }\left( \frac{\delta n_{e}}{n_{e0}}\right) +\frac{Z_{-}}{m_{-}}%
\mathbf{\nabla }U_{pond(+)}
\tag{27}
\end{equation}

On substitution of (26) into (24) and (25), one can obtain Zakharov equation for the excitation of ion-acoustic wave. 

\begin{equation}
\left( \frac{\partial ^{2}}{\partial t^{2}}-u_{0}^{2}\nabla ^{2}\right)
\left( \frac{\delta n_{-}}{n_{-0}}\right) =\frac{Z_{-}}{m_{-}}\nabla
^{2}U_{pmd(+)}
\tag{28}
\end{equation}

where $u_{0}^{2}={\frac{Z_{-}^{2}}{Z_{+}}\frac{T_{+}}{m_{-}}}$ is the ion-acoustic speed and $U_{pond(+)}=e^{2}|\mathbf{A}|^{2}/2m_{+}c^{2}$ is the ponder motive potential.  

{\bf Case 2:  Pair plasma with the concentration of electrons}

Since electrons are lighter than the ions, so here they are most effected by the wave field of \ ion sound wave via the ponderomotive force\cite{28}.  
Then the expression for the electrons density on slow timescale can be
expressed as:

\begin{equation}
n_{e}=n_{e0}\exp \left( \frac{e\phi -\frac{e^{2}\left \vert A\right \vert ^{2}%
}{2m_{e}c^{2}}}{T_{e}}\right)   \label{p2a}
\tag{29}
\end{equation}

We further suppose that $|e\varphi-e^{2}|\mathbf{A}|^{2}/2m_{e}c^{2}|\ll
T_{e}$, the potential energy $e\phi$ is given as $e\varphi=\frac{T_{e}}{n_{e0}}\left(  Z_{i}\delta n_{i+}-Z_{-}\delta
n_{-}\right)  +U_{pond(e)}$, where $U_{pond(e)}=e^{2}|\mathbf{A}|^{2}/2m_{o}c^{2}$.  From (23) and (29), we obtain

\begin{equation}
\frac{\partial \mathbf{v}_{+}}{\partial t}+\left( \mathbf{v}_{+}\cdot 
\mathbf{\nabla }\right) \mathbf{v}_{+}=-\left[ \frac{T_{e}Z_{+}}{m_{+}}%
\mathbf{\nabla }\left( \frac{\delta n_{e}}{n_{e0}}\right) +\frac{Z_{+}}{m_{+}%
}\mathbf{\nabla }U_{pond(e)}\right] 
 \tag{30}
\end{equation}

and for the negative ions. 

\begin{equation}
\frac{\partial \mathbf{v}_{-}}{\partial t}+\left( \mathbf{v}_{-}\cdot 
\mathbf{\nabla }\right) \mathbf{v}_{-}=-\frac{T_{e}Z_{-}}{m_{-}}\mathbf{%
\nabla }\left( \frac{\delta n_{e}}{n_{e0}}\right) +\frac{Z_{-}}{m_{-}}%
\mathbf{\nabla }U_{pond(e)}
\tag{31}
\end{equation}

We now assume that the ponderomotive force due to the  sound waves is not strong
enough to affect the nonlinearity of the ions, i.e., we can neglect the
second term of {the} l.h.s. in Eqs. (30-31) and linearize the continuity
equation. After some straightforward algebraic steps, we obtain Zakharov equation for the excitation of ion acoustic waves.

\begin{equation}
\left( \frac{\partial ^{2}}{\partial t^{2}}-{u}_{s}^{\ast 2}\mathbf{\nabla }%
^{2}\right) \frac{\delta n_{e}}{n_{e0}}=\left( \frac{Z_{+}^{2}n_{+0}}{m_{+}}+%
\frac{Z_{-}^{2}n_{-0}}{m_{-}}\right) \mathbf{\nabla }^{2}U_{pond(e)}. 
\tag{32}
\end{equation}

where ${u}_{s}^{\ast }=\left[ \frac{T_{e}}{m_{+}}\left( \frac{Z_{+}^{2}n_{+0}}{%
n_{e0}}\right) +\frac{T_{e}}{m_{-}}\left( \frac{Z_{-}^{2}n_{-0}}{n_{e0}}%
\right) \right] $ is the ion acoustic speed.

\section{Self focusing of acoustic wave and compression of ions}

Up to now we have discussed in Sec (II), the spectrum of ion sound wave for different cases.
Now with the aid of  Zakharov equation (32), we analyze subsonic and supersonic cases for the localization of the ion sound wave that
 will help us understand how the compression of pair ions takes place in the localized area 
of ion sound wave. 
 
 {\bf Subsonic Case:} 

First we suppose that the subsonic case for that $\partial
_{t}^{2}\ll u_{s}^{\ast 2}\mathbf{\nabla }^{2}$
so that Eq. (32) reduces to

\begin{equation}
\left( \frac{\delta n_{e}}{n_{e0}}%
\right) = - \frac {\alpha} {u_{s}^{\ast 2}} U_{pond(e)} <0. 
 \tag{33}
\end{equation}

where $\alpha =\left({Z_{+}^{2}n_{+0}}/{m_{+}}+{Z_{-}^{2}n_{-0}}/{m_{-}}\right) $. From above expression it is 
clear, when the energy density of the sound wave increases, then at that point the density of the electrons 
decreases. In other words the pair particles {(positive and negative ions)} cluster there whereas no electrons exist so in that
 region one can have possible gathering of pair ions. 
 
 {\bf Supersonic Case:} 
When the perturbations are fast that is in the supersonic regime $\partial
_{t}^{2}\gg u_{s}^{\ast 2}\mathbf{\nabla }^{2}$, Eq. (32) gives%

\begin{equation}
\frac{\partial ^{2}}{\partial t^{2}}\left( \frac{\delta n_{e}}{n_{e0}}%
\right) =\left( \frac{Z_{+}^{2}n_{+0}}{m_{+}}+\frac{Z_{-}^{2}n_{-0}}{m_{-}}%
\right) \mathbf{\nabla }^{2}U_{pond(e)}.  
\tag{34}
\end{equation}%

We now assume that the ion acoustic wave propagates along z-axis and
introduce $r_{\bot }$ and $\tau =t-z/v_{g}$ as new variables, then Eq.(34) becomes

\begin{equation}
\frac{\partial ^{2}}{\partial t^{2}}\left( \frac{\delta n_{e}}{n_{e0}}%
\right) =\frac{\alpha }{v_{g}^{2}}\left( \frac{\partial ^{2}}{\partial t^{2}}%
+\nabla _{\bot }^{2}\right) |U_{pond(e)}|^{2}.
\tag{35}
\end{equation}

After integrating twice to Eq. (35), we get%

\begin{equation}
\left( \frac{\delta n_{e}}{n_{e0}}\right) =\frac{\alpha }{v_{g}^{2}}\left \{
|U_{pond(e)}(r_{\bot },\tau )|^{2}+v_{g}^{2}\nabla _{\bot }^{2}\int_{\tau _{1}}^{\tau
}\int_{\tau _{1}}^{\tau ^{\prime }}|U_{pond(e)}(r_{\bot },\tau ^{\prime \prime })|^{2}%
\text{ }d\tau ^{\prime \prime }d\tau ^{\prime }+C_{1}\tau +C_{2}\right \}.
\tag{36}
\end{equation}

If we choose the boundary condition such that $|U_{pond(e)}|=0$
then the Eq. (36) gives $C_{1}=C_{2}=0.$ In order to find the
different feature of Eq. (35), we suppose that $|\Psi (r_{\bot },\tau
)|^{2}$ has the following profile 

\begin{equation}
U_{pond(e)}^{2}=a_{0}^{2}\exp [-r^{2}/2R^{2}]\{ \Theta (\tau -\tau
_{1})-\Theta (\tau -\tau _{2})\},
\tag{37}
\end{equation}

where $R$ is the pulse width, $\tau _{2}-\tau _{1}=\tau _{0}$ is the
pulse length and $\Theta (x)$ is a unit step function with properties \ $%
\Theta (x)=0$ for $x<0$ and $\Theta (x)=1$ for $x>0.$ Substituting (37) in (36) gives

\begin{equation}
\left( \frac{\delta n_{e}}{n_{e0}}\right) =\frac{\alpha |U_{pond(e)}(r_{\bot },\tau
)|^{2}}{v_{g}^{2}}\left[ 1-v_{g}^{2}\frac{\tau _{0}^{2}}{R^{2}}(1-\frac{%
r^{2}}{2R^{2}})a_{0}^{2}\exp (-r^{2}/2R^{2})\right].
\tag{38}
\end{equation}

For the case $r_{\bot }=0,$ and $\tau -\tau _{1}=\tau _{0},$ then ${ |U_{pond(e)}(0,\tau
)|^{2}} = a_{0}^{2}$ = constant. Then the above equation (38) implies%

\begin{equation}
\left( \frac{\delta n_{e}}{n_{e0}}\right) =\frac{\alpha a_{0}^{2}}{v_{g}^{2}}%
\left[ 1-\frac{v_{g}^{2}\tau _{0}^{2}}{R^{2}}\right].
\tag{39}
\end{equation}

Using above expression, we discuss two cases when the ion sound wave is focused.
First when the wave has a bullet like initial profile i.e. $v_{g}\tau
_{0}>R$, we get   

\begin {equation}
\left( \frac{\delta n_{e}}{n_{e0}}\right) =-\frac{\alpha a_{0}2 \tau _{0}^{2}} {
R^{2}}
\tag{40}
\end {equation}

which means in the focusing region density of the electrons decreases ($\delta n_{e}<0$) and for the depletion of electrons
there only pair particles (positive and negative ions) are left. This is in particularly important for us because as stated in the
introduction the presence of electrons (may be in small concentration) is inevitable; however, if we wish to study how the pure pair plasmas are going to 
behave then this investigation is very useful. 

One the other hand  when  the pulse shape is pancake $\left( \delta
n_{e}>0\right)$ hence there exists a regime of interaction of the wave
packet, where electrons can assemble together in the focusing region
of the wave packet. 

However, for a two component pure pair plasma, we can see the force which acts on the negative ions, for that Zakharov equation (28) follows

\begin{equation} 
\frac{\delta n_{-}}{n_{-0}}=- \frac {\beta U_{pond(+)}} {u_{0}^{2}-u_{p}^2}
\tag{41}
\end{equation}

where $\beta=Z_{-}/m_{-}$ and $u_{0}^{2}= Z_{-}^{2}{T_{+}}/(Z_{+}m_{-})$. There have introduced $u_{p}$ as the characteristic velocity of the pulse. 
From Eq. (27), we find, $F= {\partial \mathbf{v}_{-}}/{\partial t}$, where $F$ is the force acting on the negative ions which is now given as

\begin{equation} 
F=- \left(\frac {\beta u^2} {u_{0}^{2}-u_{p}^2}\right) \nabla{U_{pond(+)}} 
\tag{42}
\end{equation}

Hence at $u_{0}^2 > u_{p}^2$    the leading half of the pulse pushes the particles (negative ions) forwards and due to the smallness 
  ${\bf v_{-}} \ll  u_{p}$, the pulse overlaps this  gathering of the particles. The back half of the pulse breaks the particles  and therefore in the region of the pulse location we have an increase of the  density.

\section{Conclusions}

In this paper, we have found a new mode, namely, the acoustic mode in pair plasma both with and without the the presence of electrons. Here difference in  temperature and mass of the ions is very important and we note that damping of ions contributes significantly to the acoustic mode. Ion sound waves have been excited by the monoenergetic electron beam of small density, and the Cherenkov instability condition for the beam particles has been found in both the hydrodynamic and kinetic limits. After interaction with plasma, the ions sound mode has been shown to generate ion acoustic type waves. 

We have shown that in our case the ion sound wave can be focused and that the pair ions cluster in the focus region. Also when the ion sound wave pulse has the form of a light bullet, the pair ions  density increases in the focusing region of the EM wave, while the opposite happens when the  wave pulse has a pancake shape.
In future, we plan to investigate the modulation instability and  both stationary and non stationary solutions.  The present results may have importance in both laboratory and astrophysical plasmas where there is a possibility of  ions with different mass and temperatures.

\textbf{Acknowledgments}:
One of us (Z.E) is grateful to  Renato Fedele, Davy Tskhakaya, Hamid Saleem and Qamar Haque for the fruitful discussions.


\begin{thebibliography}{99}
\bibitem{1} W. Oohara and R. Hatakeyama, Phys. Rev. Lett. \textbf{91}, 205005 (2003). 
\bibitem{2} W. Oohara, D. Date, and R. Hatakeyama, Phys. Rev. Lett. \textbf{95}, 175003 (2005).
\bibitem{3} W. Oohara, T. Hibino, T. Higuchi, and T. Ohta, Rev. Sci. Instrum. 83, 083509 (2012).
\bibitem{4} H. Saleem Phys. Plasmas 14, 014505 (2007) and references therein, and H. Saleem, Phys. Plasmas 13, 044502 (2006) and references therein.
\bibitem{5}  F. Verheest, Phys. Plasmas 13, 082301 (2006).
\bibitem{6}  J Vranjes and S Poedts, Plasma Sources Sci. Technol 14, 485 (2005).
\bibitem{7}  M. Kono, J. Vranjes and N. Batool, Phys. Rev. Lett. 112, 105001 (2014).
\bibitem{8}  D Lu, Z. L Li , B.S. Xie , J. Plasma Physics, 81 905810508 (2015).
\bibitem{9}  F. Verheest, Phys. Plasmas  17, 062302 (2010).
\bibitem{10} Q. Haque, Physics Letters A 374, 3304Ð3307 (2010). 
\bibitem{11} S. M. Mahajan, N. L. Shatashvili and V. I. Berezhiani, Phys Rev E 80, 066404 (2009)  and references therein.
\bibitem{12} F. Pacini, Nature London 219, 145 (1968).
 \bibitem{13} P Goldreich and W. H. Julian, Astrophys. J. 157, 869 (1969).
 \bibitem{14} M. J. Rees, Nature London 229, 312 (1971).
\bibitem{15} C. M. Surko, M. Leventhal, and A. Passner, Phys. Rev. Lett. 62, 901 (1989).
\bibitem{16} H. Boehmer, M. Adams, and N. Rynn, Phys. Plasmas 2, 4369 (1995). 
\bibitem{17} E. P. Liang, S. C. Wilks, and M. Tabak, Phys. Rev. Lett. 81, 4887 (1998).
 \bibitem{18} S.A. Khan, M. Ilyas, Z. Wazir and Z. Ehsan, Astrophys Space Sci DOI 10.1007/s10509-014-1925-8.
\bibitem{19} S. Wineberg, Gravitation and Cosmology: Principles and Applications of the General Theory of Relativity (Wiley and Sons, New York, 1972).
\bibitem{20}V. I. Berezhiani, S. M. Mahajan, and N. L. Shatashvili, Phys Rev. A 81, 053812 (2010).
\bibitem{21} V. I. Berezhiani, S. M. Mahajan, and N. L. Shatashvili, J. Plasma Physics, 76, 467 (2010).
\bibitem{22} A. Esfandyari-Kalejahi, I. Kourakis, M. Mehdipoor, and P. K. Shukla, J. Phys. A: Math. Gen. 39, 13817 (2006). 
\bibitem{23} V. I. Berezhiani, and N. L. Shatashvili, , S. M. Mahajan and B N. Aleksic Phys Rev. E 88, 015101 (2013). 
 \bibitem{24} W. Oohara and R. Hatakeyama, Phys. Plasmas 14, 055704 (2007).
\bibitem{25} W. Oohara, Y. Kuwabara, and R. Hatakeyama, Phys. Rev. E 75, 056403 (2007).
\bibitem{26} V. E. Zakharov, Zh. EKSP. and Teor. Fiz. \textbf{62}, 1745 (1972).
\bibitem{27} R. Fedele, P. K. Shukla, M. Onorato, D. Anderson, and M. Lisak, Phys. Lett. A 303, 61 (2002).
\bibitem{28} N. L. Tsintsadze, Z. Ehsan, H. A. Shah and G. Murtaza, Phys. Plasmas \textbf{13}, 072103 (2006).




\end{thebibliography}
\end{document}